\documentclass[draft,jgrga]{agutex}

 \usepackage[dvips]{graphicx}
 \setkeys{Gin}{draft=false}
\authorrunninghead{RUIZ et al.}
\titlerunninghead{Aging of anisotropy in SW turbulence}

\authoraddr{M.E. Ruiz,
Instituto de Astronom\' ia y F\' isica del Espacio (CONICET-Universidad de Buenos 
Aires), C.C. 67, Sucursal 2 28, 1428, Buenos Aires, Argentina.
(meruiz@iafe.uba.ar)}

\authoraddr{S. Dasso,
Instituto de Astronom\' ia y F\' isica del Espacio (CONICET-Universidad de Buenos 
Aires) C.C. 67, Sucursal 2 28, 1428, Buenos Aires, and Departamento de F\' isica, 
Facultad de Ciencias Exactas y Naturales, Universidad de Buenos Aires, Pabellon 
1 (1428), Buenos Aires, Argentina.
(dasso@df.uba.ar)}

\authoraddr{W.H. Matthaeus,
Department of Geography, Bartol Research Institute, University of Delaware, 
Newark, DE, USA. 
(whm@udel.edu)}

\authoraddr{E. Marsch, 
Max-Planck-Institut f\"{u}r Sonnensystemforschung, Max-Planck-Stra{\ss}e 2, 
Katlenburg-Lindau, Germany.
(marsch@linmpi.mpg.de)}

\authoraddr{J.M. Weygand, 
Institute of Geophysics and Planetary Physics, University of 
California, Los Angeles, CA, USA.
(jweygand@igpp.ucla.edu)}

\begin{document}

%
%
\title{Aging of anisotropy of solar wind magnetic fluctuations
in the inner heliosphere}
%



\authors{M.E. Ruiz, 	 \altaffilmark{1}
	 S. Dasso, 	 \altaffilmark{1,2}
	 W.H. Matthaeus, \altaffilmark{3}
	 E. Marsch,	 \altaffilmark{4}
         and
	 J.M. Weygand,  \altaffilmark{5}}

\altaffiltext{1}{Instituto de Astronom\' ia y F\' isica del Espacio 
(CONICET-Universidad de Buenos Aires), Buenos Aires, Argentina.}

\altaffiltext{2}{Departamento de F\' isica, Facultad de Ciencias
Exactas y Naturales, Universidad de Buenos Aires, Buenos Aires, Argentina.}

\altaffiltext{3}{Bartol Research Institute, University of Delaware, Newark, DE, USA.}

\altaffiltext{4}{Max-Planck-Institut f\"{u}r Sonnensystemforschung, 
Max-Planck-Stra{\ss}e 2, Katlenburg-Lindau, Germany.}

\altaffiltext{5}{Institute of Geophysics and Planetary Physics, University of 
California, Los Angeles, CA, USA.}


\begin{abstract}
We analyze the evolution of the interplanetary magnetic field spatial structure
by examining the inner heliospheric autocorrelation function, using Helios
1 and Helios 2 {\it in situ} observations. 

We focus on the evolution of the integral length scale ($\lambda$) anisotropy
associated with the turbulent magnetic fluctuations,
with respect to the {\it aging} of fluid
parcels traveling away from the Sun, and according to whether the measured
$\lambda$ is principally parallel ($\lambda_\parallel$) or perpendicular
($\lambda_\perp$) to the direction of a suitably defined local ensemble 
average magnetic field $\bf{B_0}$.
We analyze a set of 1065 $24$-hour long intervals (covering full missions). 
For each interval, we compute the magnetic autocorrelation function, using
classical single-spacecraft techniques, and estimate $\lambda$ with help of two
different proxies for both Helios datasets.
We find that close to the Sun, $\lambda_{\parallel}<\lambda_{\perp}$. This
supports a slab-like spectral model, where the population of fluctuations
having wavevector $k$ parallel to $\bf{B_0}$ is much larger than the one with
$k$-vector perpendicular. A population favoring perpendicular $k$-vectors
would be considered quasi-two dimensional (2D). Moving towards $1$~AU, we find
a progressive isotropization of $\lambda$ and a trend to reach an inverted
abundance, consistent with the well-known result at $1$~AU that
$\lambda_\parallel>\lambda_\perp$, usually interpreted as a dominant 
quasi-2D picture over the slab picture.
Thus, our results are consistent with driving modes having wavevectors parallel
to $\bf{B_0}$ near Sun, and a progressive dynamical spectral transfer of energy
to modes with perpendicular wavevectors as the solar wind parcels age while
moving from the Sun to $1$ AU.

\end{abstract}

%
%

\begin{article}

%
%
\section{Introduction}

The solar wind (SW) is a natural laboratory to study magnetohydrodynamic
(MHD) turbulence, being the most completely studied case of turbulence in
astrophysics, and the only one extensively and directly studied using {\it in
situ} observations. Here we focus on the dynamical development of anisotropy
in the SW magnetic fluctuations, a well-known property of an MHD system with
a mean magnetic field ($\bf{B_0}$). Understanding the nature and origin
of this anisotropy is of relevance not only for the study of turbulence
itself, but also because magnetic fluctuations directly influence the
transport (i.e., acceleration and scattering) of solar and galactic energetic
particles in the solar wind.

A magnetized turbulent MHD system will develop anisotropies with respect to
the mean magnetic field $\bf{B_0}$. This was originally demonstrated
in laboratory experiments \citep{RobRus71,ZwebenEA79} and was also well
documented in analytical, numerical and observational studies
\citep{Shebalin83,Montgomery82,Oughton94,GS95}.
The simplest models commonly used for the description of anisotropic SW
fluctuations are the \textquotedblleft slab\textquotedblright~model, where
fluctuations have wave vectors parallel to $\bf{B_0}$, and the
\textquotedblleft 2D\textquotedblright~model, where fluctuations have wave
vectors perpendicular to $\bf{B_0}$
\citep[e.g.,][]{Matthaeus90,Oughton94,TuMarsch93,BieberEA94,BieberEA96}.
These models are of course greatly oversimplified but provide a useful
parametrization of anisotropy in SW turbulence.
In analytical calculations a two component decomposition of this type, having
no oblique wave vectors, provides great simplifcations, for example in
scattering and transport theory
\citep[e.g.,][]{MatthaeusEA95,ShalchiEA04-parperp}. However, conceptually it
is often advantageous to think of the two components as \textquotedblleft
quasi-slab\textquotedblright~and \textquotedblleft
quasi-2D\textquotedblright, meaning that according to some specified scheme
based on time scales, angle, etc., all wavevector contributions are grouped
into these two categories. This is the approach we adopt here, and hereafter
we shall refer to the two relevant populations as simply slab or 2D
components.

Anisotropies have been widely investigated at $1$ astronomical unit (AU) for 
more than $20$ years, using single spacecraft techniques
\citep[e.g.,][]{BelcherDavis71,Matthaeus90, TuMarsch1995,Milano2004,
Dasso05}, that means, analyzing under the Taylor frozen-in hypothesis
\citep{Taylor1938}, spatial structures from time series measured by one
spacecraft. Recently, multi spacecraft studies at $1$ AU have validated the
main results on anisotropic turbulence obtained from single spacecraft
observations
\citep{MatthaeusET05,Dasso08,Weygand09,Osman2007Apj,MatthaeusET10}.

As a consequence of these numerous studies, a common assumption is that the
SW at $1$ AU, contains a major population of 2D fluctuations and a minor slab
component. However, single spacecraft studies have shown that, when
subdividing the sample into fast SW (speeds larger than $500$ km/s) and slow
SW (speeds lower than $400$ km/s), the former contains more slab-like than
2D-like fluctuations, while in the latter it is the other way round
\citep{Dasso05}. This result has recently been confirmed by means of
multi spacecraft techniques \citep{Weygand11}, where the slow SW was defined
as having speeds below $450$ km/s and the fast SW above $600$ km/s.

Thus, all these results at $1$ AU motivate the following questions: Is the
distinct relative population of fluctuations in fast and slow SW streams a
consequence of intrinsic differences in the fluctuation properties that are
established at the coronal sources? Or, if we assume that a high-speed stream
will arrive at $1$ AU more quickly than a low-speed stream, thus revealing
\textquotedblleft younger\textquotedblright~states in the evolution of
interplanetary turbulence, are these differences a consequence of the
dynamical evolution of the turbulence from the Sun to $1$ AU? In the latter
case, this issue becomes relevant to questions of dynamical evolution of
inner heliospheric turbulence in a more general way.

It has long been known that the shape of the interplanetary magnetic field
energy spectrum evolves \citep[e.g.,][]{BavassanoEA82,TuEA84,RobertsEA87a,
RobertsEA87b, Marsch90} radially at Helios orbital distances between $0.3$ and
$1$ AU. This includes migration of the \textquotedblleft bendover
scale\textquotedblright~that separates a Kolmogorov-like $\sim -5/3$ spectral
slope at higher frequencies from a flatter $\sim 1/f$ spectral form at lower
frequencies, which is associated with the observed increase of the magnetic 
autocorrelation length for growing helidistances (e.g.,
\cite{BrunoEA1986,RuizSW12}).
Similarly, there is a transition from statistically more pure
Alfv\'enic fluctuations close to the Sun to more mixed, but still
predominantly outward, fluctuations farther away from the Sun, where they become 
a mixture of \textquotedblleft inward\textquotedblright~and \textquotedblleft 
outward\textquotedblright~type Els\"{a}sser fluctuations, and then tend to become 
equipartitioned near $1$ AU \citep{Marsch90} and beyond. 
These radial changes have been attributed to dynamical evolution of
Alfv\'enic turbulence.
Other authors, \cite[e.g.,][]{Bavassano1989}, have suggested that solar wind 
fluctuations may be a superposition of convected structures (of probably solar 
origing) and propagating and nonlinearly evolving Alfv\'en waves. 

On the other hand, there have been opposite suggestions that certain features
of turbulence do not evolve with heliocentric distance \citep{MacBrideEA10},
or even that the majority of the spectral power in SW fluctuations is not
active turbulence at all \citep{Borovsky08,Borovsky10}.

To answer these questions better, it is desirable to understand more fully
how (or if) anisotropy evolves with distance from the Sun in the inner
heliosphere, thereby taking into account (to the extent this is possible) the
dynamical age of the turbulence as well as the magnetic field direction and
the SW speed. 
We address these questions in the present work employing a wide range of 
plasma-parcel ages ranging from 20 up to 140 hours. The results show that
the anisotropy measured by the correlation scale systematically
evolves in the inner heliosphere, indicating the presence of a
rich set of dynamical processes to be explored by planned
missions such as Solar Probe Plus and Solar Orbiter.

In the following section we present the theoretical background for our work.
In section \ref{OBS} we describe the data processing, while in section
\ref{ANYS} we analyze the anisotropy of the correlation lengths. Finally, in
section \ref{CONCLU} we present a discussion of the results and our
conclusions.

\section{Magnetic autocorrelation function}\label{MCF_}

A turbulent magnetic field $\bf B$ can be studied by separating the
small-scale fluctuating component ${\bf b}$ from the large-scale field $\langle
{\bf B} \rangle$, described by an ensemble average (indicated by brackets
$\langle \dots \rangle $). Thus the field is decomposed as:
\begin{equation}
  {\bf B}=\langle  {\bf B} \rangle + {\bf b} 
\end{equation}
The examination of the magnetic self-correlation function, which can be
defined as
\begin{equation}  
R(\{{\bf x}, t\};\{{\bf r}, \tau\}) = \langle {\bf b}({\bf x}, t) \cdot {\bf
b}({\bf x} + {\bf r}, t + \tau) \rangle , \label{R_def}
\end{equation}
is one way of analyzing turbulent magnetic field fluctuations. $R(\{{\bf x},
t\};\{{\bf r}, \tau\})$ represents the average trace of the usual
two-points/two-times correlation tensor for the magnetic field, and $\{{\bf
r},\tau\}$ the space and time lags, respectively.
The dependence on the position and time $\{{\bf x},$t$\}$, where $R$ is
computed, can be removed if we suppose a stationary and homogeneous medium.
Since the SW (with radial velocity ${\bf V_{sw}} = \hat{r}V_{sw}$) is
supersonic and super-Alfv\'enic with respect to the spacecraft (considered to
be at rest in the heliosphere during the several hours of measurement), we
may assume the validity of the Taylor frozen-in-flow hypothesis
\citep{Taylor1938},
\begin{equation}  
R({\bf r}=-\hat{r}V_{sw} \tau,\tau=0) \Bigl |_{fluid \ frame} = R({\bf
r=0},\tau) \Bigl |_{spacecraft \ frame} \label{R_taylor_1},
\end{equation}
that is, fluctuations are just convected past the spacecraft in a time scale
shorter than its own characteristic dynamical time scale. Then, the intrinsic
time dependence of the magnetic fluctuations in Equation \ref{R_def} can be
neglected, and $R$ becomes a function of ${\bf r}$ alone. Therefore the
spatial structure of $R$ can be computed from the time series of the field
observed {\it in situ} by the spacecraft and reads:
\begin{equation}  
R({\bf r}) = \langle {\bf b}({\bf 0}) \cdot {\bf b}({\bf r}) \rangle
\label{R_taylor}.
\end{equation}
Suppose we consider a spatial lag, ${\bf r}(\theta) = r \hat{\bf r}(\theta)$,
that lies in the direction $\hat {\bf r}$ making an angle $\theta$ to the mean magnetic field ${\bf B_{0}}$. For isotropic turbulence, correlations fall
off in the same way in any direction. However, for anisotropic turbulence, the
correlation function will not behave the same way in all directions. A
measure of this correlation anisotropy can be constructed by computing an
integral scale from the correlation function. This may also be called the
spatial correlation length along the direction given by $\theta$, and can be
defined as
\begin{equation}
\lambda (\theta) = \frac{ \int^{\infty}_{0} \langle {\bf b}({\bf 0}) \cdot
{\bf b}(r \hat {\bf r}(\theta) \rangle d r}{<b^2>} \label{lambda_def},
\end{equation}
where $\hat {\bf r} (\theta)$ is a unit vector defining the direction of
integration.
Following convention, this correlation length $\lambda (\theta)$ can be viewed
as an anisotropic measure of the size of the energy-containing eddies in turbulence
\citep{Batchelor70}.
The correlation scale is also frequently regarded as a demarcation of the
low-frequency end of the power-law range, separating the inertial range from
the low-frequency spectrum that is associated with large-scale structures in
the SW \citep{LivRev05}.

\section{Data Analysis}\label{OBS}

In this section we apply the theoretical approach summarized above to the
data sets obtained from the magnetic field \citep{Neubauer77Hmagneto} and
plasma \citep{Rosen77plasma,Marsch82Hprotons} instruments onboard the Helios
1 (H1) and Helios 2 (H2) spacecrafts. The time series we analyze correspond
to the {\it in situ} solar wind observations made from December 1974 to June
1981 (comprising almost one full eleven-year solar cycle). They have a
cadence of $40$~s and are essentially on the ecliptic plane between
$0.3$ AU and $1.0$ AU.
We group the data into $24$-hour intervals ($I$), thus obtaining $N_{1}$
subseries (or intervals, $I=\{ 1,...,N_{1} \}$). Then we repeat this
procedure by shifting the data by $12$ hours to obtain $N_{2}$ additional
intervals. This approach maximizes utilization of the data.
To avoid samples with very low statistical significance, we retain only those
intervals encompassing at least the $30 \%$ of the observations expected for
the cadence mentioned above. Then, our collection of
usable data includes $N_{1}+N_{2}=N=705$ intervals for spacecraft H1 and
$N=743$ intervals for spacecraft H2.

The correlation functions and associated correlation scales are computed in
the following manner. From the observed time series of the magnetic field
(${\bf B}^{I}$), we define in each interval the magnetic fluctuations as
${\bf b}^{I}={\bf B}^{I}-{\bf B}^{I}_{0}$, where ${\bf B}^{I}_{0}$ is a
linear fit to ${\bf B}^{I}$ data. This procedure removes both the mean
value of the field and a linear trend associated with unresolved very
low-frequency power. The local estimate of the ensemble average magnetic
field is identified with the fit field ${\bf B}^{I}_{0}$.

Then, we employ the {\it Blackman-Tukey} technique to compute each
correlation function $R^{I}$ in the same way as done in \cite{Milano2004}.
To be able to compare intervals with different fluctuation amplitudes, we
compute normalized correlation functions as $R({\bf r})^{norm,I}=R({\bf
r})^{I}/R({\bf 0})^{I}$. For simplicity of notation, we omit the ``norm''
label hereafter.

\begin{figure}
\noindent\includegraphics[width=20pc]{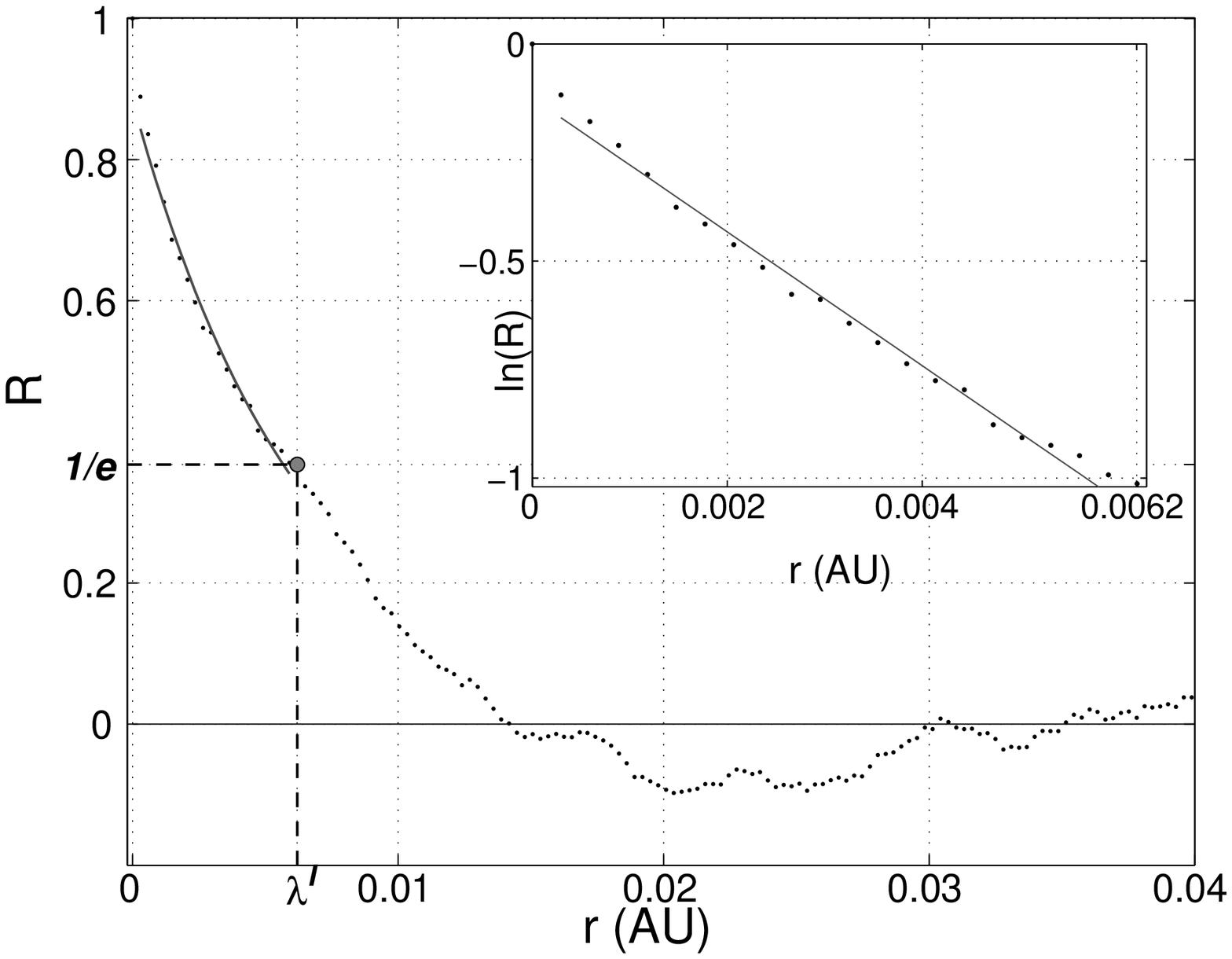}
\caption{Typical (variance-normalized) magnetic self-correlation function $R$
         in the inner heliosphere. This single interval corresponds to Helios 1
         observations at $0.964$ AU on December 26, 1974.
         The inset shows $R$ in logarithmic scale. Solid line shows the
         linear fit of $\ln(R)$ vs. $r$.}
\label{R_tip}
\end{figure}

A typical magnetic correlation function, computed by the above method in the 
inner heliosphere, is shown in Figure \ref{R_tip}.
We use two different methods, called {\it i} and {\it ii}, to estimate the
magnetic correlation length ($\lambda^{I}$) from the correlation function in
each interval. A simple approximation for the behavior of $R^{I}$ at the
large scales and in the long-wavelength part of the inertial range is an
exponential decay, $R \sim e^{-r/\lambda}$.
Method {\it i} determines an estimate of the correlation scale $\lambda^{I}_{i}$
as the value of the lag $r$ where the decreasing function $R^{I}$ reaches
$\exp(-1)$ by first time, that is $R^{I}(\lambda^{i}_i) = 1/e$.
Using the same approximation for the form of the correlation, we can
parameterize the correlation function as $\log (R) \sim -r/\lambda$. The
second method {\it ii} employs this relation and defines $\lambda^{I}_{ii}$
as minus the inverse of the slope obtained from a linear fit to $\log (R)$
vs. $r$ (see inset plot in Figure \ref{R_tip}).

\begin{figure}
\noindent\includegraphics[width=20pc]{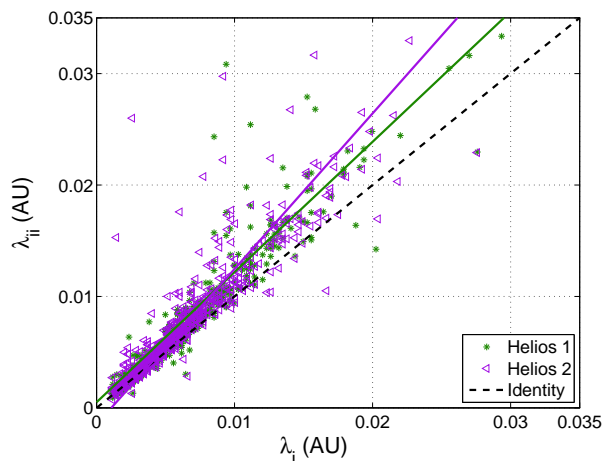}
\caption{Scatter plot of $\lambda_{ii}^{I}$ vs. $\lambda_{i}^{I}$ for Helios 1
         and Helios 2. The line $\lambda_{ii}^{I} = \lambda_{i}^{I}$ and linear
        least-squares lines are shown for reference.}
\label{compare_L}
\end{figure}

A comparison of methods {\it i} and {\it ii} is shown in Figure
\ref{compare_L}. On average $\lambda^{I}_{ii}$ provides an
estimate that is slightly larger than the one provided by $\lambda^{I}_{i}$,
$\lambda^{I}_{ii} / \lambda^{I}_{i} \sim 1.16$, as revealed by
least-squares fits to the Helios data.

We also include in our data set the mean within each interval of the distance
from the Sun to the spacecraft ($D^{I}$), the proton speed
($V_{p}^{I}=V_{sw}^{I}$), and the angle ($\theta^{I}$) between the direction
of the mean magnetic field $\langle {\bf B}^{I}_{0} \rangle$ and the SW
velocity (${\hat {\bf r}}$, which gives the direction of the spatial lag).

As a final step, we refine the analyzed data set by removing outliers of
$\lambda^{I}$. Values of $\lambda^{I}$ are considered as spurious if they
depart from their means by one standard deviation $\sigma$ toward smaller
values of $\lambda$, or by $2\sigma$ toward larger ones. 
Also considered outliers are those
intervals in which the mean-field direction is not well determined. We select
intervals with the direction of $\langle{\bf B}^{I}_{0} \rangle$ (and
therefore $\theta^{I}$) well established inside the 24-hour interval.
Accordingly, we define the quantity $\Delta\theta$ as the difference between
the mean angle $\theta$ in the first half and second half of the interval;
and retain only those intervals showing low values of $\Delta\theta$,
specifically $|\Delta\theta|<30^{\circ}$.
After the above refinements, the final data set encompasses $536$ intervals
for H1 and $529$ for H2, totaling $1065$ intervals. Although there are several gaps
in the Helios data set, the final $1065$ intervals are roughly equi-distributed
during the solar cycle and along the spacecraft orbit. 

\section{Anisotropy}\label{ANYS}

We proceed to analyze the dependence of $\lambda$ upon heliodistance $D$, angle 
relative to $\langle {\bf B}^{I}_{0} \rangle$, and the age of the turbulence.

We defined three heliodistance ranges (bins), namely, close to the Sun,
intermediate distances, and near $1$ AU, each bin having a width of $0.23$ AU
($0.22$ AU) for H1 (H2). We established two angular channels: the parallel
channel ($0^{\circ}<\theta^{I}<40^{\circ}$) and the perpendicular one
($50^{\circ}<\theta^{I}<90^{\circ}$). 
Narrower angular bins were explored, arriving at 
qualitatively similar results but
with larger errorbars (lower number of intervals per bin). Thus, we conclude
that widening the angular channels up to $40^{\circ}$ increases the parallel
and perpendicular counts without polluting the results. Nevertheless, smaller 
angular bins have been used in other works when studying power and spectral
anisotropies \citep[e.g.,][]{Horbury2008PRL,Wicks2010MNRAS}
using methods that differ from what we employ here.

We then distribute all the intervals
into the spatial bins and angular channels and compute conditional averages
of $\lambda$. In this way, for every heliocentric distance bin we obtain
eight mean values of $\lambda$: four $\lambda_{\parallel}$ and four
$\lambda_{\perp}$. The factor of four is associated with all combinations of
the two spacecraft and the two methods.

The relative order between $\lambda_{\parallel}$ and $\lambda_{\perp}$ can be
interpreted qualitatively in terms of the relative abundance of the two basic
components, quasi-slab and quasi-2D, of the MHD-scale fluctuations.
Accordingly, preponderance of the slab-like component with wavevectors mainly
parallel to the mean magnetic field $\langle {\bf B}^{I}_{0} \rangle$ is identified 
by $\lambda_{\parallel}/\lambda_{\perp} < 1$, while preponderance of 
the quasi-2D component, having mainly perpendicular wavevectors, is identified by
$\lambda_{\parallel}/\lambda_{\perp}>1$ \citep{Matthaeus90,Dasso05}.

\begin{figure}
\noindent\includegraphics[width=120mm,height=60mm]{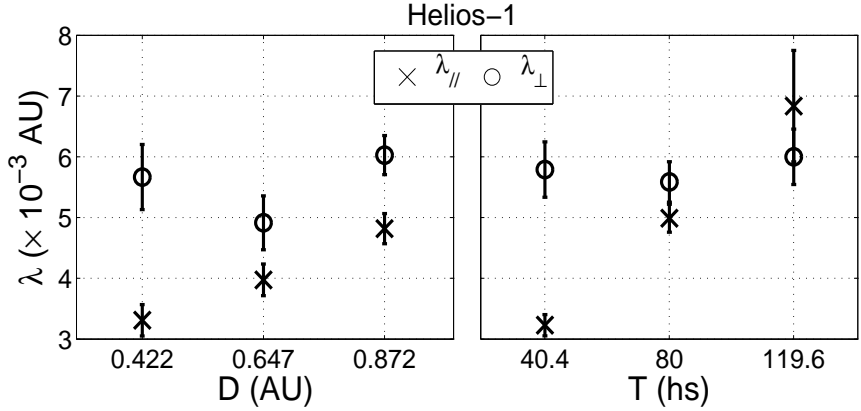}
\caption{Evolution of $\lambda_{i,\parallel}$ and $\lambda_{i,\perp}$.
         Left panel: dependence on {\it heliodistance} ($D$).
         Right panel: dependence on {\it turbulence age} ($T$).
         Ticks of the abscissas indicate the center of each bin,
         bars show the mean's error.}
\label{ANI_AGE_DIST}
\end{figure}

The left panel in Figure \ref{ANI_AGE_DIST} shows the evolution of
$\lambda_{\parallel}$ and $\lambda_{\perp}$ with the heliocentric distance
for the H1 spacecraft. The data plotted correspond to correlation lengths
computed with method $i$ (we obtain similar results with method $ii$ and both
methods for H2, not shown here for brevity).

For the group of datasets at distances nearest the Sun, we find that the
turbulence is highly anisotropic, with $\lambda_{\parallel} /
\lambda_{\perp}=0.58, 0.51, 0.89, 0.81$ for method-spacecraft combinations
$i$-H1, $ii$-H1, $i$-H2, $ii$-H2, respectively. As the heliocentric distance
grows, the observed anisotropy becomes weaker, due to a steadily growth
of $\lambda_{\parallel}$ while $\lambda_{\perp}$ remains nearly constant,
implying a shift of $k_{\parallel}$ towards small frequencies, as has been
previously reported \citep{LivRev05}.
Several previous works based
on observations at $1$ AU have shown that $\lambda_{\parallel} >
\lambda_{\perp}$ only for slow SW, as well as for samples with mixed fast and
slow SW \citep{Matthaeus90,Milano2004,Dasso05,Weygand09}.

It must be noted that, near the ecliptic plane, the slow wind is much more
frequent than the fast wind, and therefore all mean values computed for a
mixed SW sample will favor slow SW properties and disfavor fast SW features.
In spite of the latter, in the region nearest $1$ AU (ranging from $0.76$ to
$0.98$ AU for H1 and from $0.75$ to $0.98$ AU for H2), we find that
$\lambda_{\parallel} < \lambda_{\perp}$.

Motivated on result at $1$ AU that shows different relative order between
$\lambda_{\parallel}$ and $\lambda_{\perp}$ for fast and slow SW
and considering that the high-speed wind will
arrive earlier at $1$ AU (and therefore being younger than the slow SW), we
analyze the dependence of $\lambda$ on what we call the ``age'' of the
interval. To this end, we compute this age, $T^I=D^I/V^I_{sw}$, for
each interval as the nominal time it takes a SW parcel moving at speed
${V^I_{sw}}$ to travel a given distance from the Sun to the spacecraft
located at $D^I$.

Accordingly, we defined three ranges of turbulence age in bins having
widths of $20.5$ hours ($19.5$ hours) for H1 (H2). The bin labelled $T_{1}$
is the youngest bin, centered at $40.5$ hours. Bin $T_{2}$ is at
intermediate ages centered at $80$ hours, while bin $T_{3}$ contains the
oldest samples centered at $120$ hours. In each bin we established the same
two angular channels. We compute again conditional averages of $\lambda$,
considering only those intervals ${I}$ that correspond to a given angular
channel and a given age range, in the same way as described above.

The results taking into account turbulence age, using spacecraft H1 and
method $i$, are shown in right panel of Figure~\ref{ANI_AGE_DIST}. Similar
results have been obtained using method $ii$ and H2 (not shown). For ages near 
$40$ hours, that is for the young wind in bin $T_{1}$, we find
that $\lambda_{\parallel} \sim \lambda_{\perp}/2$. In particular
$\lambda_{\parallel}/\lambda_{\perp}=0.56,0.91,0.48,0.61$, for
method-spacecraft $i$-H1, $i$-H2, $ii$-H1, $ii$-H2. This strong anisotropy is
consistent with the injection of $k$-parallel Alfv\'en waves near the Sun. As
the wind age grows towards bin $T_{2}$, around the age of $T=80$
hours, we find a trend to an isotropization of $\lambda$. For ages
between $T=80$ hours and $T=120$ hours an inversion happens, and when the
wind is already old that is for ages about $120$ hours in bin
$T_{3}$, we find that $\lambda_{\parallel}>\lambda_{\perp}$, as expected at $1$
AU for a slow or mixed SW. Values of $\lambda_{\parallel} / \lambda_{\perp}$
in each aging bin for the different combinations of spacecraft and
method are shown in Table~\ref{tbl-2}, showing robust results.

\begin{figure}
\noindent\includegraphics[width=30pc,height=20pc]{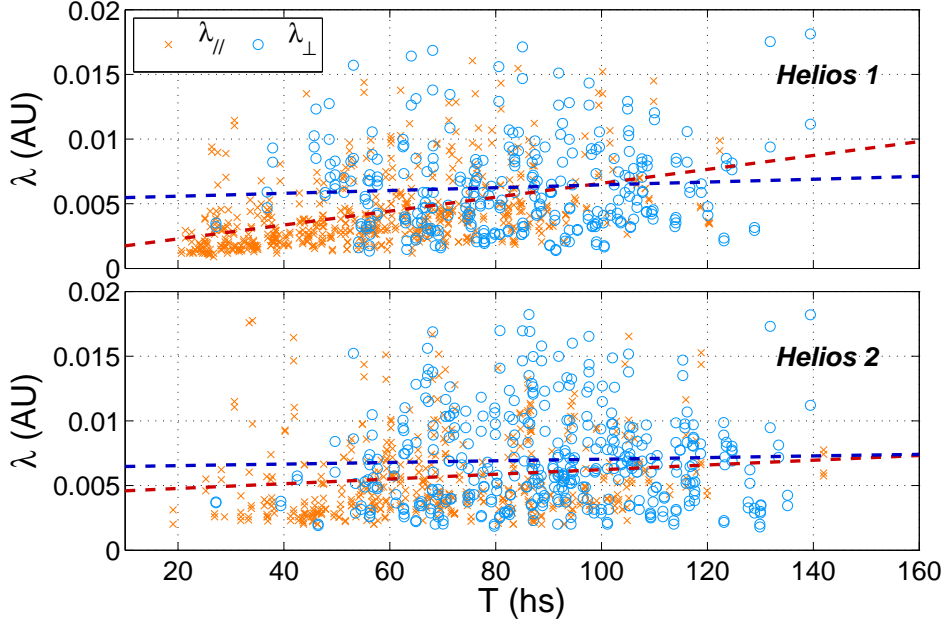}
\caption{Scatter plot of $\lambda_{\parallel}$ and $\lambda_{\perp}$
         vs. age. Here both methods $i$ and $ii$ have been considered.
         Linear trends are shown for reference:
         Red line is fit for $\lambda_{\parallel}$;
         blue line for $\lambda_{\perp}$.}
\label{cool_plot}
\end{figure}

Figure~\ref{cool_plot} presents as a function of age, and without any
binning, the values of $\lambda_{\parallel}$ and $\lambda_{\perp}$ obtained
for each interval. Single lines showing linear fits to data (red line for
$\lambda_{\parallel}$ and blue line for $\lambda_{\perp}$) reveal a tendency
of $\lambda_{\parallel}$ to grow with $T$ at a significantly larger rate than
$\lambda_{\perp}$. In fact, with both H1 and H2 datasets it is possible to
find a linear behavior of $\lambda_{\parallel}$, which can be represented as
$\lambda_{\parallel, s/c}(T)= a_{\parallel, s/c}T + b_{\parallel, s/c}$,
where $s/c$ is H1 or H2. Least-squares fits give the coefficients:
$a_{\parallel, H1}=(5.3 \pm 0.5) \times 10^{-5}$ AU h$^{-1}$, $a_{\parallel,
H2}=(1.8 \pm 0.7) \times 10^{-5}$ AU h$^{-1}$, $b_{\parallel, H1}=(1.2 \pm
0.3) \times 10^{-3}$ AU and, $b_{\parallel, H2}=(4.4 \pm 0.5) \times 10^{-3}$
AU.

In contrast, for $\lambda_{\perp}$, if there is a linear increase, it is
apparently so weak that neither H1 nor H2 could observe it, since $a_{\perp,
H1}=(1 \pm 1) \times 10^{-5}$ AU hs$^{-1}$ and $a_{\perp, H2}=(0.6 \pm 0.9)
\times 10^{-5}$ AU hs$^{-1}$, i.e., within error bars the slope may be zero.
Nevertheless, a dimensionless global estimator for the evolution of the
anisotropy in the range $10 < T < 200$ hours may be given by the relation:
$Q(T)=\lambda_{\parallel}/\lambda_{\perp} = (AT +B)/\langle \lambda_{\perp}
\rangle$. Averaging the data from both missions we obtain $A=(3.6 \pm 1.7)
\times 10^{-5}$ AU hs$^{-1}$, $B=(2.8 \pm 1.6) \times 10^{-3}$ AU (assigned
errors correspond to the semi difference of the Helios values) and $\langle
\lambda_{\perp} \rangle= 0.007 \pm 0.004 $ AU.

\section{Discussion and conclusions}\label{CONCLU}

From the analysis of almost seven years of Helios SW data, we have presented
a study of the dynamical evolution of the anisotropy of magnetic fluctuations
in the inner heliosphere. The main indicator we employ in this paper for the
study of SW anisotropy is the correlation scale measured either mainly
parallel to, or mainly perpendicular to, the mean magnetic field direction.

Helios 1 and Helios 2 are unique spacecraft that systematically have {\it in
situ} observed the inner heliosphere, and therefore they can up to this day
provide the best available evidence of the turbulence state close to the
Sun, and give hints on the possible initial conditions of the turbulence in
the outer corona. The here derived measures of the correlation-scale
anisotropy also provide indications relevant for planning the {\it in situ}
observations by the upcoming {\it Solar Probe Plus} and {\it Solar Orbiter}
missions.

Especially in the inner heliosphere, the state of the turbulence, including
the nature of the anisotropies studied here, is a complex combination of
boundary and source characteristics. They are related to the initiation of
the SW and convolved with various types of evolution that may commence very
close to the Sun, perhaps at the Alfv\'enic critical point (possibly located
within 14-34 $R_s$; see e.g. \cite{Marsch84}).
It appears, however, that the observed evolution of the anisotropy with
heliocentric distance, as found in this study, is in agreement with several
turbulence theories that favor perpendicular spectral transfer of energy.
Generally, one expects from experiment, theory and numerical simulations that
the anisotropic cascade preferentially magnifies gradients perpendicular to
the mean field, while the evolution of the wavevector component parallel to
the local mean magnetic field is relatively suppressed
\citep[e.g.,][]{zweben81,Shebalin83,RobRus71,Oughton94,GS95}.

The results presented here are consistent with a greater abundance of the
slab-like population near Sun, which progressively evolves, according to the
SW age, which is defined as the time spent by fluid parcels since their birth
near the Sun until the {\it in situ} observation. During this evolution, the
relative abundance of the slab-like and 2D populations tends to become
inverted, and the system reaches at $1$ AU a state resembling that of the
slow wind which has a stronger 2D component. 

%
In this work we study the spatial structure of the magnetic correlation
function, detecting strong anisotropy near the Sun and weak anisotropy near 1
AU.
Power and spectral index anisotropies have also been studied by other authors 
\citep[e.g.,][]{Horbury2008PRL,Podesta2009ApJ,Wicks2010MNRAS,Narita_PRL_2010,
Narita_JGR_2010}. These authors have employed methods that effectively define
the mean field to be a local quantity that itself depends upon the fluctuations.
This way of looking at the anisotropy (based on the local mean field) is
very different from the approach used in the present work. Here we treat the
mean field as an estimate of an ensemble average quantity, an approach which
favors maintaining well defined ensemble averaged directions (i.e., a fixed
coordinate system for evaluation of the spectra and correlation functions).
Although both methods differ, we view them to be complementary. Examples
of the local perspective are given by
\cite{Cho_2000ApJ,Maron_sim_2001ApJ,Horbury2008PRL,Podesta2009ApJ,
Wicks2010MNRAS}. A comparison between the local and global methods is shown
by \cite{Chen_2011MNRAS}.
In particular, \cite{Wicks2010MNRAS} using $100$ continuous days of Ulysses data
reported isotropy of the outer scale for fast polar wind in the outer 
heliosphere between $1.38$ and $1.93$ AU. On the other hand,
\cite{Narita_PRL_2010,Narita_JGR_2010}, using $4$ hours of Cluster data
analyzed the energy distribution of magnetic field fluctuations in the $3$D wave
vector domain and obtained direct evidence of the dominance of the 2D picture.

In this manner, our results are fully compatible with earlier findings that the
fast SW (which is younger) is more slab-like and that the slow SW (which is
older) is more 2D-like \citep{Dasso05,Weygand11}.

These results have direct impact on diffusion models for transport of charged
particles through the inner heliosphere, indicating that the energy distribution
near the Sun is richer in slab modes than in 2D modes.

Here we have focused on the anisotropy of SW magnetic fluctuations, our
future studies will revise the velocity as well as Els\"{a}sser-variable
fluctuations.

\begin{acknowledgments}
SD is member of the Carrera del Investigador Cientifico, CONICET. MER is a
fellow from CONICET. MER and SD acknowledge partial support by Argentinean
grants: UBACyT 20020090100264 (UBA), PIP 11220090100825/10 (CONICET), PICT
2007-00856 (ANCPyT). SD acknowledges support from the Abdus Salam International
Centre for Theoretical Physics (ICTP), as provided in the frame of his regular
associateship. JW and WHM acknowledge partial support by NASA Heliophysics
Guest Investigator Program grant NNX09AG31G, and NSF grants ATM-0752135 (SHINE) 
and ATM-0752135.
\end{acknowledgments}
\end{article}



%
\newpage

\begin{table}
\caption{Ratios of $\lambda_{\parallel}$ to $\lambda_{\perp}$ for 
        each age bin ($T^{I}$)\label{tbl-2}}
\begin{tabular}{ccccc}
\tableline\tableline
Spacecraft & Method & $\frac{\lambda_{\parallel}}{\lambda_{\perp}}$ ($T_{1}$)
                    & $\frac{\lambda_{\parallel}}{\lambda_{\perp}}$ ($T_{2}$)
                    & $\frac{\lambda_{\parallel}}{\lambda_{\perp}}$ ($T_{3}$)\\   
\tableline
Helios 1 & $i$  & $0.56 \pm 0.05$ & $0.89 \pm 0.07$ & $1.14 \pm 0.18$\\
         & $ii$ & $0.48 \pm 0.06$ & $0.77 \pm 0.07$ & $1.01 \pm 0.16$\\
\tableline
Helios 2 & $i$  & $0.91 \pm 0.11$ & $0.84 \pm 0.05$ & $0.99 \pm 0.10$\\
         & $ii$ & $0.61 \pm 0.11$ & $0.80 \pm 0.06$ & $0.99 \pm 0.11$\\
\tableline
\tableline
\end{tabular}
\end{table}

\end{document}